&pdflatex

\documentclass[prl,twocolumn,aps,amssymb,amsmath,tightenlines,showpacs]{revtex4}
\usepackage{graphicx,epsfig}

\def\half{{\textstyle{\frac{1}{2}}}}
\def\cP{\mathcal P}
\def\cC{\mathcal C}
\def\cT{\mathcal T}
\def\cPT{\mathcal PT}

\begin{document}
% \preprint[{\rightline{KCL-PH-TH/2015-{\bf XX}, LCTS/2015-15}}
% \vspace{1cm}

\title{$\cPT$-symmetric interpretation of unstable effective potentials}

\author{Carl M. Bender$^{a,b}$}\email{cmb@wustl.edu}
\author{Daniel W. Hook$^{a,c}$}\email{d.hook@imperial.ac.uk}
\author{Nick E. Mavromatos$^{b,d}$}\email{nikolaos.mavromatos@kcl.ac.uk}
\author{Sarben Sarkar$^b$}\email{sarben.sarkar@kcl.ac.uk}

\affiliation{$^a$Department of Physics, Washington University, St. Louis, MO
63130, USA\\
$^b$Theoretical Particle Physics and Cosmology Group, King's~College~London,
London WC2R 2LS, UK\\
% $^c$Department of Mathematical Science, City University London,\\
% Northampton Square, London EC1V 0HB, UK\\
$^c$Theoretical Physics, Imperial College London, London SW7 2AZ, UK\\
$^d$Theory Division, CERN, CH-1211 Geneva 23, Switzerland}

\begin{abstract}
The conventional interpretation of the one-loop effective potentials of the
Higgs field in the Standard Model and the gravitino condensate in dynamically
broken supergravity is that these theories are unstable at large field values. A
$\cPT$-symmetric reinterpretation of these models at a quantum-mechanical level
eliminates these instabilities and suggests that these instabilities may also be
tamed at the quantum-field-theory level.
\end{abstract}

\date{\today}
\maketitle

The use of effective potentials to study symmetry breaking~\cite{R1,R2} in field
theory has a long history and much is known about the structure of such
effective theories. In the case of a four-dimensional conformally invariant
theory of a scalar field $\varphi$ interacting with fermions and gauge fields,
the renormalized effective potential has the form $\Gamma[\varphi]=\varphi^4f
\left[\log\left(\varphi^2/\mu^2\right),g\right]$, where $\mu$ is a mass scale
and $g$ denotes the coupling constants in the theory \cite{R3,R4}. Different
theories are distinguished by the function $f$. The large-$\varphi$ behavior of
the effective action determines the stability of the vacuum state.

Here, we consider two theories of current interest. The first is a theory of
dynamical breaking of gravity via a gravitino condensate field $\varphi$
\cite{R5,R6}. For this case
\begin{equation}
\label{E1}
\Gamma[\varphi]\propto-\varphi^4\log(i\varphi)
\end{equation}
for large $\varphi$. The second is the Standard Model of particle physics for
which $\varphi$ is the Higgs field and for this case \cite{R7}
\begin{equation}
\label{E2}
\Gamma[\varphi]\propto-\varphi^4\log(\varphi^2)
\end{equation}
for large $\varphi$. Evidently, radiative corrections and renormalization can
lead to effective potentials that suggest that the theory is unstable (and that
it has complex energy levels).

An early observation that renormalization can cause instability was made by
K\"all\'en and Pauli \cite{R8}, who showed that upon renormalizing the Lee model
\cite{R9} the Hamiltonian becomes complex, ghost states arise, and the S-matrix
becomes nonunitary. However, a $\cPT$-symmetric analysis tames the apparent
instabilities of the Lee model: energies are real, ghost states disappear, and
the S-matrix becomes unitary \cite{R10}. $\cPT$-symmetric quantum theory also
repairs the ghost problem in the Pais-Uhlenbeck model \cite{R11}, the illusory
instability of the double-scaling limit of O($N$)-symmetric $\varphi^4$ theory
\cite{R12,R13}, and difficulties associated with the complex Hamiltonian for
timelike Liouville field theory \cite{R14}.

This Letter examines three $\cPT$-symmetric quantum-mechanical Hamiltonians
associated with the two problematic quantum field theories above. ($\cP$ denotes
parity reflection $x\to-x$, $p\to-p$; $\cT$ denotes time reversal $x\to x$, $p
\to-p$, $i\to-i$ \cite{R15}.) The first Hamiltonian,
\begin{equation}
H=p^2+x^4\log(ix),
\label{E3}
\end{equation}
is a toy model to study logarithmic $\cPT$-symmetric theories. We show that the
spectrum of this complex $\cPT$-symmetric Hamiltonian is discrete, real, and
positive. The second Hamiltonian,
\begin{equation}
H=p^2-x^4\log(ix),
\label{E4}
\end{equation}
is the quantum-mechanical analog of (\ref{E1}). We show that the spectrum of
this complex and apparently unstable Hamiltonian is also discrete, real, and
positive, and this suggests that there is no instability in the supergravity
theory of inflation in Ref.~\cite{R6}. The third Hamiltonian,
\begin{equation}
H=p^2-x^4\log(x^2),
\label{E5}
\end{equation}
is motivated by the renormalized effective potential for Higgs model (\ref{E2}).
We show that the ground-state energy of this Hamiltonian is real and positive,
and this suggests the intriguing possibility that, contrary to earlier work
\cite{R16}, the Higgs vacuum may be stable.

These three models all have a new $\cPT$-symmetric structure that has not
previously been examined, namely, the logarithm term in the Hamiltonian. We show
that the $\cPT$ symmetry of the Hamiltonians (\ref{E3}) and
(\ref{E4}) is {\it unbroken}; that is, their spectra are entirely real. However,
the $\cPT$ symmetry of $H$ in (\ref{E5}) is {\it broken}; only the four
lowest-lying states have real energy. Thus, while the ground state is stable,
almost all other states in the theory are unstable. This is indeed what is
observed in nature; almost all particles are unstable and there are only a few
stable particles. This suggests the conjecture that the Higgs vacuum is stable
as a consequence of $\cPT$ symmetry and that the universe may be described by a
Hamiltonian having a broken $\cPT$-symmetry.

\vspace{.1cm}

\noindent{\bf Analysis of the toy model Hamiltonian (\ref{E3}):} To analyze $H$
in (\ref{E3}) we first locate the complex turning points. Next, we examine the
complex classical trajectories on an infinite-sheeted Riemann surface and find
that all these trajectories are closed. This shows that the energy levels are
all real \cite{R17}. Last, we perform a WKB calculation of the eigenvalues and
note that the results agree with a precise numerical determination of the
eigenvalues.

The turning points for $H$ in (\ref{E3}) satisfy the equation
\begin{equation}
E=x^4\log(ix),
\label{E6}
\end{equation}
where $E$ is the energy. We take $E=1.24909$ because this is the numerical value
of the ground-state energy obtained by solving the Schr\"odinger equation for
the potential $x^4\log(ix)$ (see Table~\ref{t1}).

One turning point lies on the negative imaginary-$x$ axis. To find this point we
set $x=-ir$ ($r>0$) and obtain the algebraic equation $E=r^4\log r$. Solving
this equation by using Newton's method, we find that the turning point lies at
$x=-1.39316i$. To find the other turning points we seek solutions to (\ref{E6})
in polar form $x=re^{i\theta}$ ($r>0,\,\theta\,{\rm real}$). Substituting for
$x$ in (\ref{E6}) and taking the imaginary part, we obtain
\begin{equation}
\log r=-(2k\pi+\theta+\pi/2)\cos(4\theta)/\sin(4\theta),
\label{E7}
\end{equation}
where $k$ is the sheet number in the Riemann surface of the logarithm. (We
choose the branch cut to lie on the positive-imaginary axis.) Using (\ref{E7}),
we simplify the real part of (\ref{E6}) to 
\begin{equation}
E=-r^4(2k\pi+\theta+\pi/2)/\sin(4\theta).
\label{E8}
\end{equation}
We then use (\ref{E7}) to eliminate $r$ from (\ref{E8}) and use Newton's method
to determine $\theta$. For $k=0$ and $E=1.24909$, two $\cPT$-symmetric
(left-right symmetric) pairs of turning points lie at $\pm0.93803-0.38530i$ and
at $\pm0.32807+0.75353i$. For $k=1$ and $E=1.24909$ there is a turning point at
$-0.53838+0.23100i$; the $\cPT$-symmetric image of this turning point lies on
sheet $k=-1$ at $0.53838+0.23100i$.

\begin{figure}[t!]
\begin{center}
\includegraphics[scale=0.23]{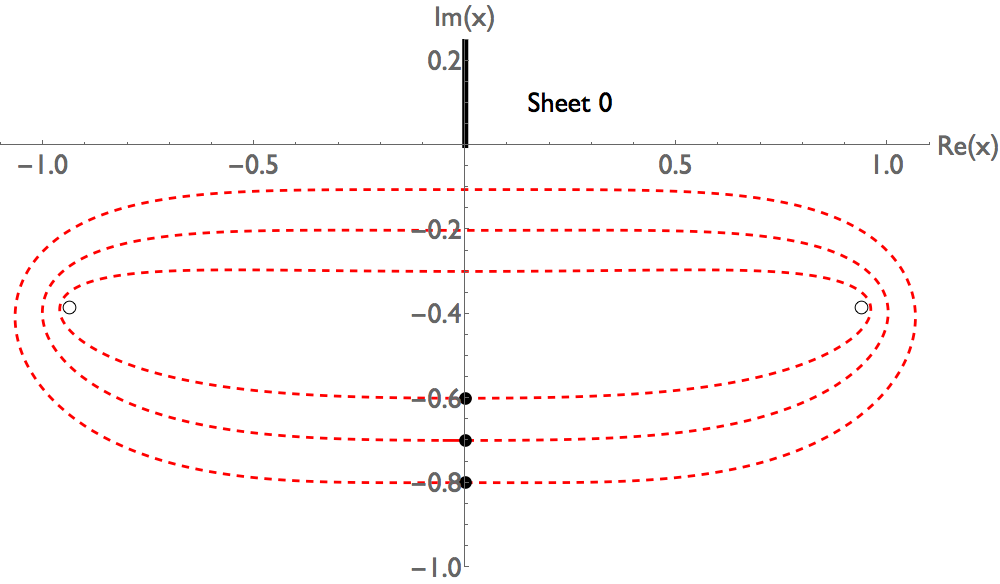}
\end{center}
\caption{[Color online] Three nested closed classical paths $x(t)$ (red dashed
curves) on the principal sheet (sheet 0) of the complex-$x$ plane for energy $E=
1.24909$ and initial conditions $x(0)=-0.6i,\,-0.7i,\,-0.8i$ (indicated by
black dots). The paths do not cross the branch cut on the positive-imaginary
axis (solid black line) so they remain on sheet 0. The paths are $\cPT$
symmetric (left-right symmetric). Turning points at $\pm 0.938-0.385i$ (small
circles) cause the paths to turn around in the right-half and left-half plane.}
\label{F1}
\end{figure}

\begin{figure}[h!]
\begin{center}
\includegraphics[scale=0.24]{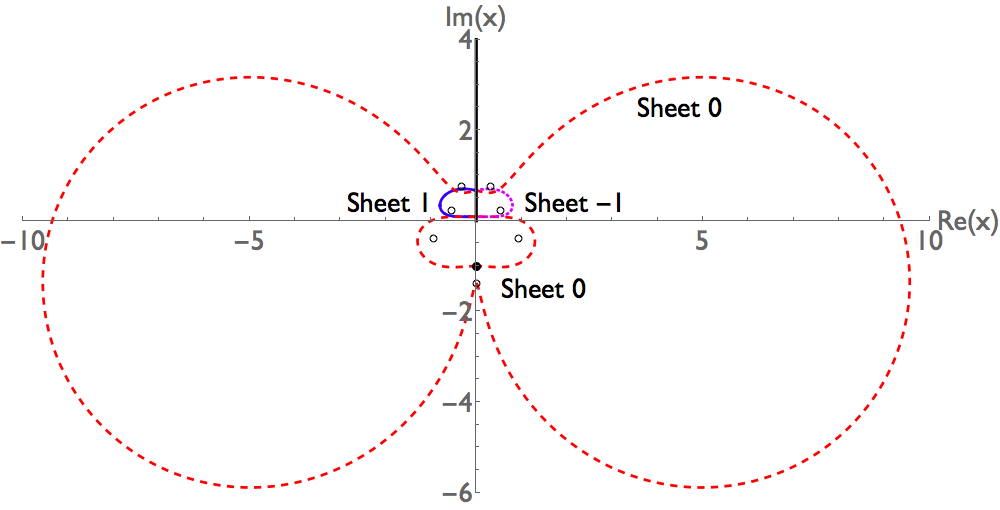}
\end{center}
\caption{[Color online] {\it Closed} classical path for energy $E=1.24909$.
This path visits three sheets in the Riemann surface. Turning points on sheets
$-1$, $0$, and $1$ determine the shape of the path. The path begins on sheet 0
at $x=-i$ at the black dot and moves to the right as a red dashed curve. It is
pulled around by the turning point (small circle) at $x=0.938-0.385i$. The path
then crosses the branch cut on the positive-imaginary axis (solid black line) at
a point just slightly above the origin and goes onto sheet 1. On sheet 1 (solid
blue line) it is pulled around by the turning point at $x=-0.538+0.231i$,
recrosses the branch cut, and goes back onto sheet 0. The path (again, a red
dashed curve) is at first under the influence of a turning point at $0.328+
0.754i$ before it is pulled around by the turning point at $x=0.938-0.385i$ and
makes a large circular loop in the right-half complex-$x$ plane. It is then
yanked around sharply by the turning point on the imaginary axis at $x=-1.393i$
and makes a $\cPT$-symmetric mirror-image loop in the left-half plane under the
influence of the turning point at $x=-0.938-0.385i$. The path then crosses the
branch cut onto sheet -1, where it becomes a dotted purple line. Finally, the
dotted path is pulled around by the turning point at $x=0.538+0.231i$ and
crosses back onto sheet 0, where it loops and returns to the initial point at
$x=-i$.}
\label{F2}
\end{figure}

The turning points determine the shape of the classical trajectories. Two
topologically different kinds of classical paths are shown in Figs.~\ref{F1} and
\ref{F2}. All classical trajectories are {\it closed} and {\it left-right
symmetric}, and this implies that the quantum energies are all real \cite{R17}.

The WKB quantization condition is a complex path integral on the principal sheet
of the logarithm ($k=0$). On this sheet a branch cut runs from the origin to
$+i\infty$ on the imaginary axis; this choice of branch cut respects the $\cPT$
symmetry of the configuration. The integration path goes from the left turning
point $x_{\rm L}$ to the right turning point $x_{\rm R}$ \cite{R18}:
\begin{equation}
\left(n+\half\right)\pi\sim\int_{x_{\rm L}}^{x_{\rm R}}dx\sqrt{E-V(x)}\quad
(n>>1).
\label{E9}
\end{equation}

If the energy is large $\left(E_n\gg1\right)$, then from (\ref{E7}) with $k=0$
we find that the turning points lie slightly below the real axis at $x_{\rm R}
=re^{i\theta}$ and at $x_{\rm L}=re^{-\pi i-i\theta}$ with
\begin{equation}
\theta\sim-\pi/(8\log r)\quad{\rm and}\quad r^4\log r\sim E.
\label{E10}
\end{equation}
We choose the path of integration in (\ref{E9}) to have a constant imaginary
part so that the path is a horizontal line from $x_{\rm L}$ to $x_{\rm R}$.
Since $E$ is large, $r$ is large and thus $\theta$ is small. We obtain the
simplified approximate quantization condition
\begin{equation}
\left(n+\half\right)\pi\sim r^3\log r\int_{-1}^1 dt\sqrt{1-t^4},
\label{E11}
\end{equation}
which leads to the WKB approximation for $n\gg1$:
\begin{equation}
\frac{E_n}{[\log(E_n)]^{1/3}}\sim\left[\frac{\Gamma(7/4)(n+1/2)\sqrt{\pi}}
{\Gamma(5/4)\sqrt{2}}\right]^{4/3}.
\label{E12}
\end{equation}
To test the accuracy of (\ref{E12}) we have computed numerically the first 14
eigenvalues by solving the Schr\"odinger equation for (\ref{E3}). These
eigenvalues are listed in Table \ref{t1}. Note that the accuracy of this WKB
approximation increases smoothly with increasing $n$.

\begin{table}[h!]
\begin{center}
\begin{tabular}{|c|c|c|c|c|}\hline
$n$ & Numerical      & $\frac{E_n}{[\log(E_n)]^{1/3}}$ & WKB        & \% error\\
    & value of $E_n$ &                                 & prediction & \\ \hline
0  & 1.24909 & 2.06161 & 0.54627 & $73.5028\,\%$ \\
1  & 4.47086 & 3.90775 & 2.36356 & $39.5161\,\%$ \\
2  & 8.76298 & 6.76804 & 4.67052 & $30.9915\,\%$ \\
3  & 13.7383 & 9.96525 & 7.31480 & $26.5969\,\%$ \\
4  & 19.2641 & 13.4195 & 10.2265 & $23.7936\,\%$ \\
5  & 25.2586 & 17.0888 & 13.3638 & $21.7983\,\%$ \\
6  & 31.6658 & 20.9458 & 16.6979 & $20.2804\,\%$ \\
7  & 38.4444 & 24.9708 & 20.2082 & $19.0730\,\%$ \\
8  & 45.5623 & 29.1487 & 23.8783 & $18.0811\,\%$ \\
9  & 52.9939 & 33.4674 & 27.6956 & $17.2463\,\%$ \\
10 & 60.7180 & 37.9172 & 31.6493 & $16.5304\,\%$ \\
11 & 68.7167 & 42.4896 & 35.7308 & $15.9070\,\%$ \\
12 & 76.9748 & 47.1776 & 39.9324 & $15.3573\,\%$ \\
13 & 85.4789 & 51.9751 & 44.2477 & $14.8676\,\%$ \\
\hline
\end{tabular}
\end{center}
\caption{\label{t1} Eigenvalues of the Hamiltonian in (\ref{E3}) compared with
the WKB approximation in (\ref{E12}). The accuracy of the WKB formula increases
smoothly with $n$.}
\end{table}

\vspace{.1cm}

\begin{figure}[h!]
\begin{center}
\includegraphics[scale=0.23]{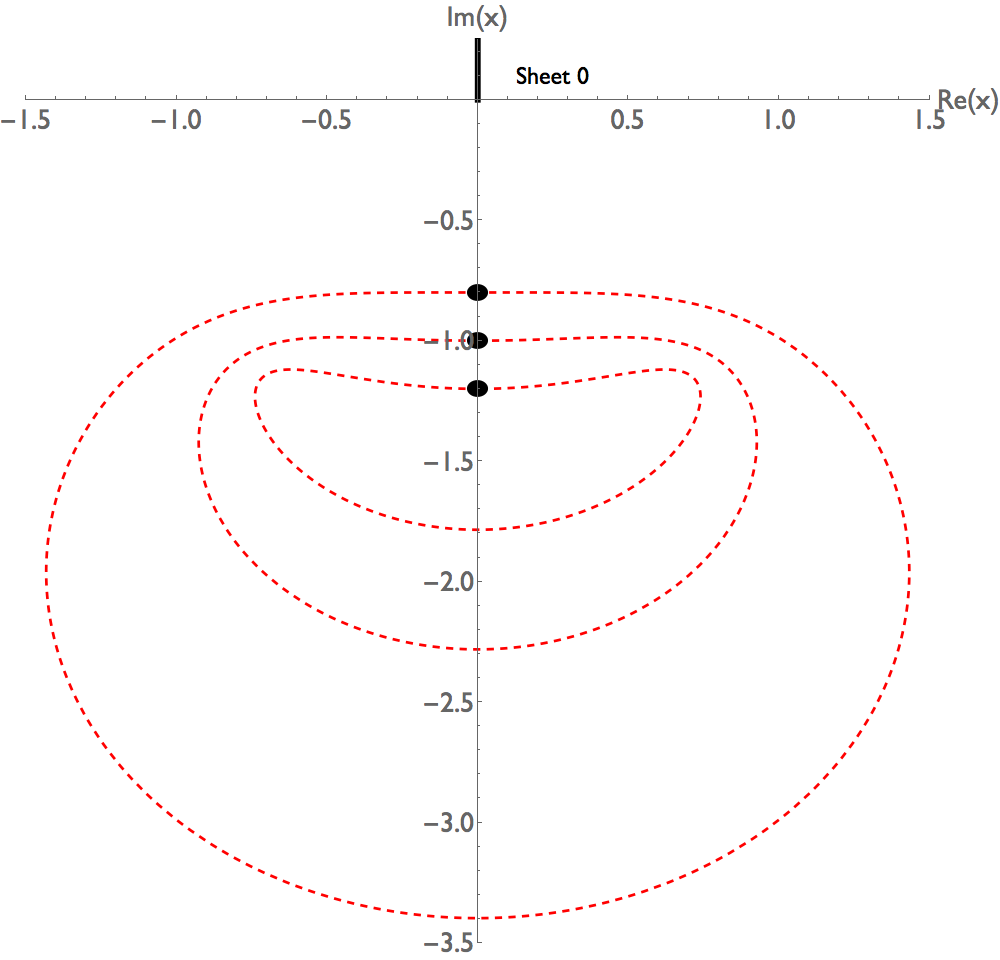}
\end{center}
\caption{[Color online] Three nested classical trajectories for $H$ in
(\ref{E4}) with $E=2.07734$. The trajectories begin at the points $-0.8i,\,-i,\,
-1.2i$ (black dots) and are closed. The trajectories do not cross the branch cut
on the positive imaginary axis (solid black line).}
\label{F3}
\end{figure}

\begin{figure}[h!]
\begin{center}
\includegraphics[scale=0.23]{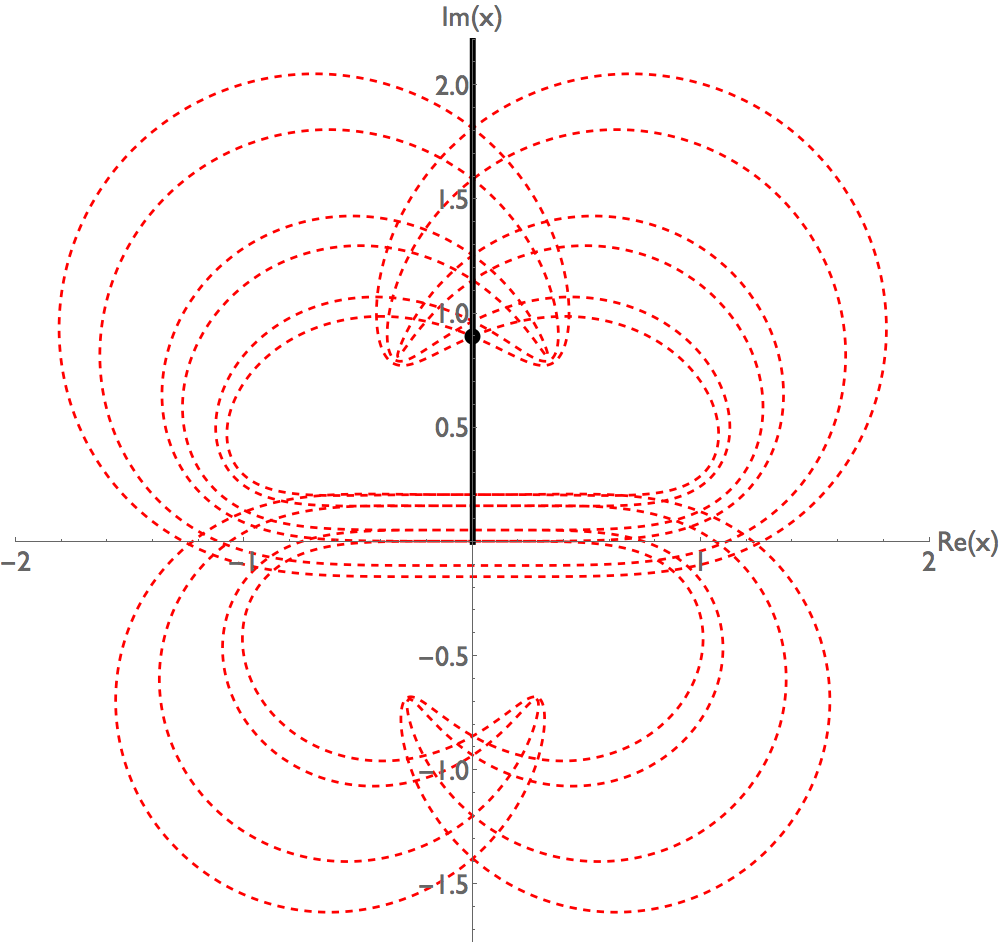}
\end{center}
\caption{[Color online] Complex classical trajectory for $H$ in (\ref{E4}) with
$E=2.07734$. The trajectory begins at $0.9i$, crosses the branch cut on the
positive-imaginary axis (solid black line), and visits three sheets of the
Riemann surface but it is still closed and $\cPT$ symmetric (left-right
symmetric).}
\label{F4}
\end{figure}

\noindent{\bf Analysis of the supergravity model Hamiltonian (\ref{E4}):}
The classical trajectories for the Hamiltonian (\ref{E4}) are plotted in
Figs.~\ref{F3} and \ref{F4}. Like the classical trajectories for the Hamiltonian
(\ref{E3}), these trajectories are closed, which implies that all the
eigenvalues for $H$ in (\ref{E4}) are real.

\begin{figure}[b!]
\begin{center}
\includegraphics[scale=0.19]{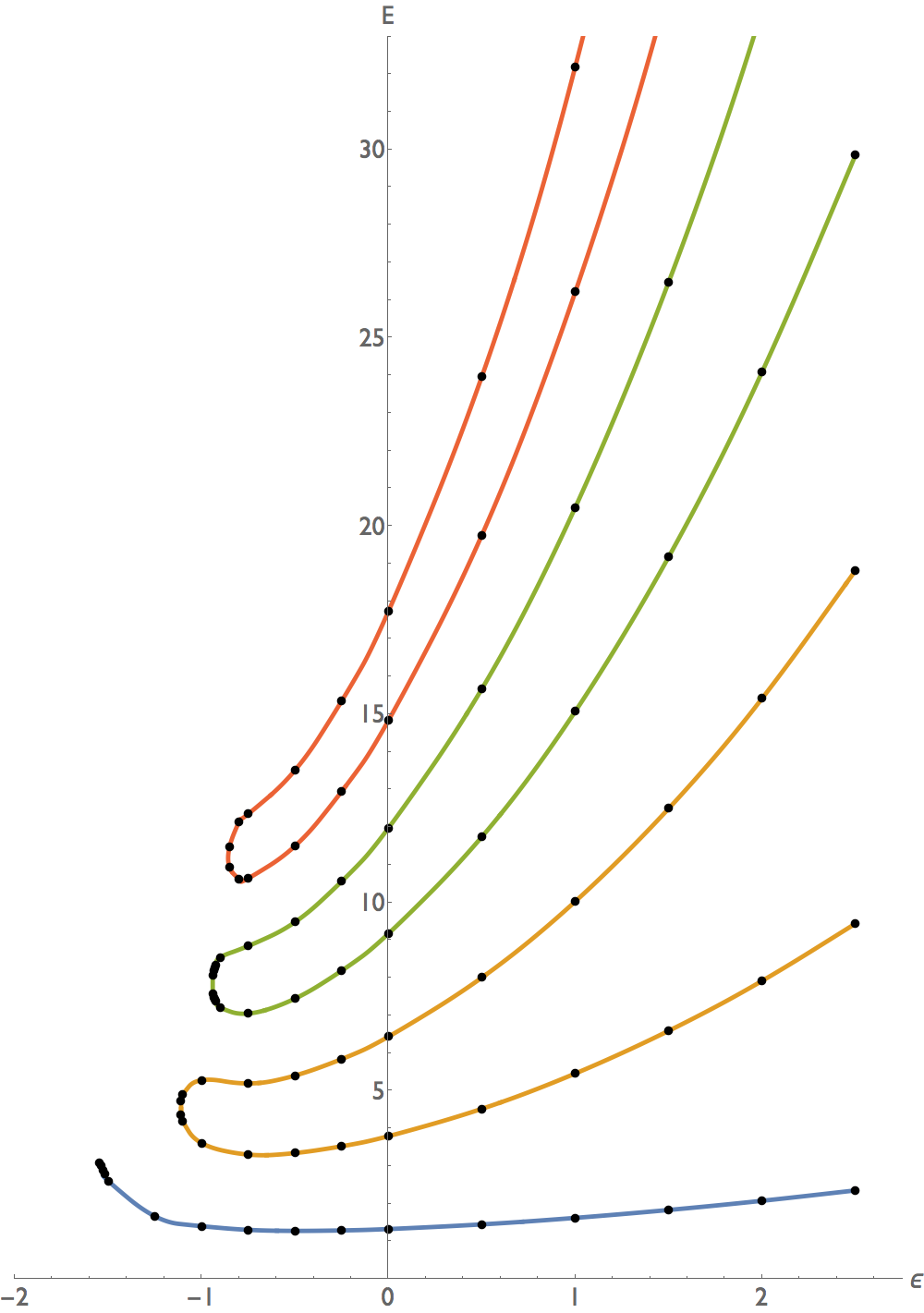}
\end{center}
\caption{[Color online] Energies of the Hamiltonian $H=p^2+x^2(ix)^\epsilon
\log(ix)$ plotted versus $\epsilon$. This Hamiltonian reduces to $H$ in
(\ref{E4}) when $\epsilon=2$. The energies are real when $\epsilon\geq0$.}
\label{F5}
\end{figure}

To find the eigenvalues of the complex Hamiltonian (\ref{E4}) we follow the
procedure described in Ref.~\cite{R15}; to wit, we obtain (\ref{E4}) as the
parametric limit $\epsilon:\,0\to2$ of the Hamiltonian $H=p^2+x^2(ix)^\epsilon
\log(ix)$, which has real positive eigenvalues when $\epsilon=0$. As $\epsilon
\to2$, the Stokes wedges for the time-independent Schr\"odinger eigenvalue
problem rotate into the complex-$x$ plane \cite{R15}. Thus, this procedure
defines the eigenvalue problem for $H$ in (\ref{E4}) and specifies the energy
levels. In Fig.~\ref{F5} we plot the eigenvalues as functions of $\epsilon$.
Note that this figure is topologically identical to Fig.~1 in Ref.~\cite{R15}
except that the ground-state energy diverges at $\epsilon=-2$ rather than at
$\epsilon=-1$ (see Ref.~\cite{R19}). This plot indicates that when $\epsilon<0$
the $\cPT$ symmetry is broken, but that when $\epsilon\geq0$ the $\cPT$ symmetry
is unbroken (all real eigenvalues).

WKB theory gives a good approximation to the eigenvalues of $H$ in (\ref{E4}).
We seek turning points for $H$ in polar form $x=re^{i\theta}$ and find that on
the principal sheet of the Riemann surface a $\cPT$-symmetric pair of turning
points lie at $\theta=-\pi/4-\delta$ and $\theta=-3\pi/4+\delta$. When $E\gg1$,
$\delta$ is small, $\delta\sim\pi/(16\log r)$, and $r$ is large, $r^4\log r\sim
E$. The WKB calculation yields a formula for the eigenvalues that is identical
to (\ref{E12}) except that there is no factor of $\sqrt{2}$ in the denominator.
Thus, for large $n$ the $n$th eigenvalue of $H$ in (\ref{E4}) agrees
approximately with the $n$th eigenvalue of $H$ in (\ref{E3}) multiplied by
$2^{2/3}$. A numerical determination of the first six eigenvalues gives
$2.07734$, $7.9189$, $15.4216$, $24.0932$, $33.7053$, and $44.1189$.

\vspace{.1cm}

\begin{figure}[t!]
\begin{center}
\includegraphics[scale=0.19]{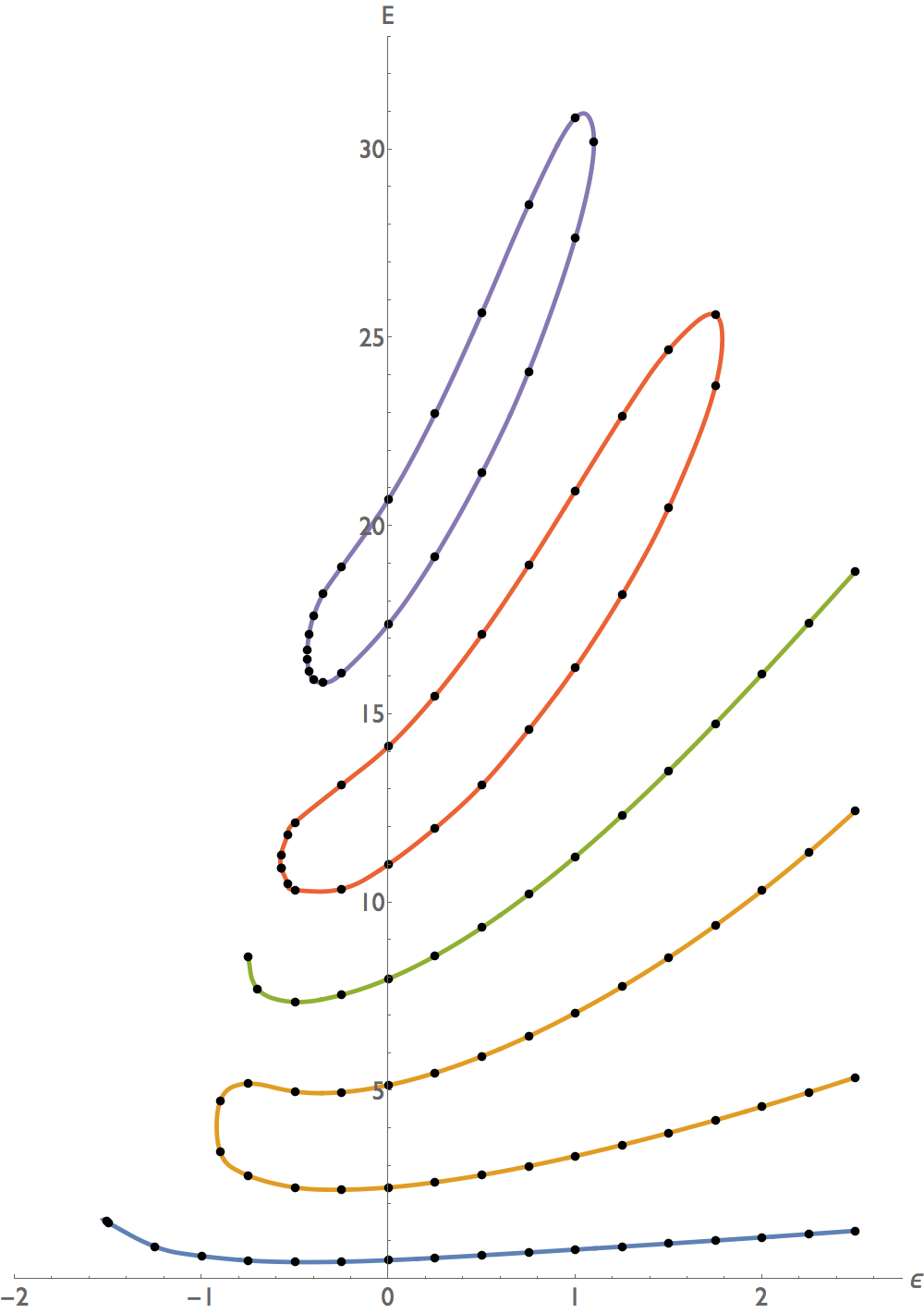}
\end{center}
\caption{[Color online] Eigenvalues of the Hamiltonian $H=p^2+x^2(ix)^\epsilon
\log\left(x^2\right)$, which reduces to $H$ in (\ref{E5}) when $\epsilon=2$.
There are four real energies when $\epsilon=2$.}
\label{F6}
\end{figure}

\begin{figure}[t!]
\begin{center}
\includegraphics[scale=0.23]{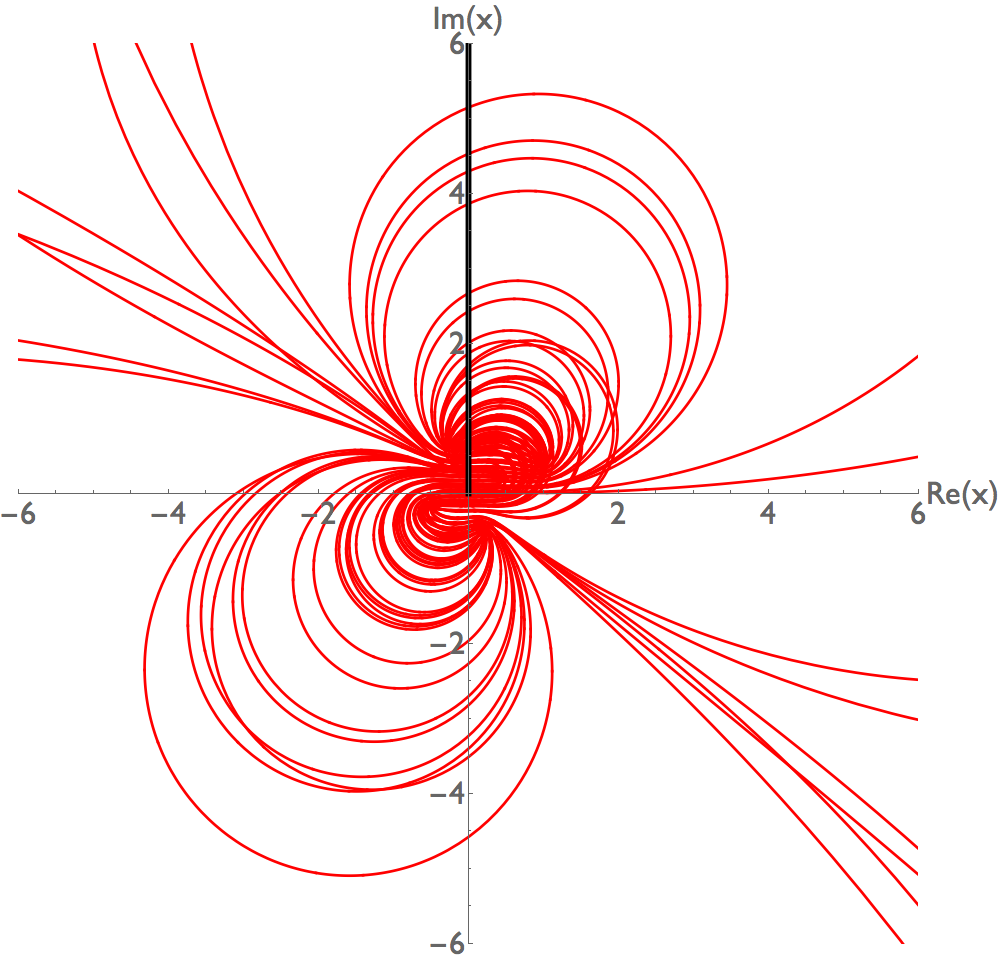}
\end{center}
\caption{[Color online] Classical paths for the Hamiltonian $H=p^2-x^4\log\left(
x^2\right)$. The initial point is $0.9i$ and the energy is $E=1.10543$. The
trajectory is not $\cPT$ symmetric. It makes bigger and bigger loops and does
not close.}
\label{F7}
\end{figure}

\noindent{\bf Analysis of the Higgs model Hamiltonian (\ref{E5}):} To make sense
of the Hamiltonian (\ref{E5}) we again introduce a parameter $\epsilon$ and we
define $H$ in (\ref{E5}) as the limit of $H=p^2+x^2(ix)^\epsilon\log\left(x^2
\right)$ as $\epsilon:\,0\to2$. This case is distinctly different from that for
$H$ in (\ref{E4}). As is illustrated in Fig.~\ref{F6}, the $\cPT$ symmetry is
broken for all $\epsilon\neq0$. Indeed, when $\epsilon=2$, there are only four
real eigenvalues: $E_0=1.1054311$, $E_1=4.577736$, $E_2=10.318036$, and $E_3=
16.06707$. To confirm this result we plot the classical trajectories for the
case $\epsilon=2$ in Fig.~\ref{F7}. In contrast with Fig.~\ref{F4} these
trajectories are not closed and are not left-right symmetric.

This result suggests that the Higgs vacuum is stable and that perhaps the real
world is in a broken $\cPT$-symmetric regime. This possibility has interesting
implications for the $\cC$ operator in $\cPT$-symmetric quantum theory. In an
unbroken regime the $\cC$ operator, which is used to construct the Hilbert-space
metric with respect to which the Hamiltonian is selfadjoint, commutes with the
Hamiltonian and thus it cannot serve as the charge-conjugation operator in
particle physics. However, in a broken $\cPT$ regime, the states of $H$ are not
states of $\cC$, and thus $\cC$ may play the role of charge conjugation in
particle physics \cite{R20}.

\vspace{.1cm}

The work of CMB and SS was supported by a Royal Society U.K. Travel Grant and
that of NEM was supported in part by the London Centre for Terauniverse Studies
(LCTS), using funding from the European Research Council via the Advanced 
Investigator Grant 267352 and by STFC (UK) under the research grant
ST/L000326/1.


\begin{thebibliography}{100}

\bibitem{R1} S.~R.~Coleman and E.~J.~Weinberg, Phys.~Rev.~D {\bf 7}, 1888
(1973).
% ``Radiative Corrections as the Origin of Spontaneous Symmetry Breaking,''
%%CITATION = PHRVA,D7,1888;%%
%3377 citations counted in INSPIRE as of 24 Apr 2015

\bibitem{R2} R.~Jackiw, Phys.~Rev.~D {\bf 9}, 1686 (1974).
% ``Functional evaluation of the effective potential,''
%%CITATION = PHRVA,D9,1686;%%
%800 citations counted in INSPIRE as of 24 Apr 2015

\bibitem{R3} K.~A.~Meissner and H.~Nicolai, Phys.~Lett.~B {\bf 648}, 312 (2007).
% ``Conformal Symmetry and the Standard Model,''
% [hep-th/0612165].
%%CITATION = HEP-TH/0612165;%%
%158 citations counted in INSPIRE as of 24 Apr 2015

\bibitem{R4} K.~A.~Meissner and H.~Nicolai, Acta Phys.~Polon.~B {\bf 40}, 2737
(2009).
% ``Renormalization Group and Effective Potential in Classically Conformal
% Theories,'' [arXiv:0809.1338 [hep-th]]
%%CITATION = ARXIV:0809.1338;%%
%16 citations counted in INSPIRE as of 24 Apr 2015

\bibitem{R5} I.~L.~Buchbinder and S.~D.~Odintsov, Class. Quant. Grav.~{\bf 6},
1955 (1989).

\bibitem{R6} J.~Alexandre, N.~Houston and N.~E.~Mavromatos, Phys.~Rev.~D {\bf
88}, 125017 (2013) and Int.~J.~Mod.~Phys.~D {\bf 24}, 1541004 (2015).
% ``Inflation via Gravitino Condensation in Dynamically Broken Supergravity,''
% [arXiv:1409.3183 [gr-qc]].
%%CITATION = ARXIV:1409.3183;%%
%5 citations counted in INSPIRE as of 24 Apr 2015

\bibitem{R7} M.~Sher, Phys.~Rept.~{\bf 179}, 273 (1989).
% ``Electroweak Higgs Potentials and Vacuum Stability,'' 273-418
%%CITATION = PRPLC,179,273;%%
%751 citations counted in INSPIRE as of 24 Apr 2015

\bibitem{R8} G.~K\"all\'en and W.~Pauli, Mat.-Fys.~Medd.~{\bf 30}, No.~7 (1955).
% ``On the mathematical structure of TD Lee's model of a renormalizable
% field theory,''

\bibitem{R9} T.~D.~Lee, Phys.~Rev.~{\bf 95}, 1329 (1954).
% ``Some Special Examples in Renormalizable Field Theory''

\bibitem{R10} C.~M.~Bender, S.~F.~Brandt, J.-H.~Chen, and Q.~Wang,
Phys.~Rev.~D {\bf 71}, 025014 (2005).
% ``Ghost Busting: $\cal{PT}$-Symmetric Interpretation of the Lee Model,''

\bibitem{R11} C.~M.~Bender and P.~D.~Mannheim, Phys.~Rev.~Lett.~{\bf 100},
110402 (2008).
% ``No-ghost Theorem for the Fourth-Order Derivative Pais-Uhlenbeck Oscillator
% Model,'' [arXiv: hep-th/0706.0207]

\bibitem{R12} C.~M.~Bender, M.~Moshe, and S.~Sarkar, J.~Phys.~A:
Math.~Theor.~{\bf 46}, 102002 (2013).
% ``PT-Symmetric Interpretation of Double-Scaling'' [arXiv: hep-th/1206.4943].

\bibitem{R13} C.~M.~Bender and S.~Sarkar, J.~Phys.~A: Math.~Theor.~{\bf 46},
442001 (2013).
% ``Double-Scaling Limit of the O(N)-Symmetric Anharmonic Oscillator,''
% [arXiv: hep-th/1307.4348].

\bibitem{R14} C.~M.~Bender, D.~W.~Hook, N.~E.~Mavromatos, and S.~Sarkar,
Phys.~Rev.~Lett.~{\bf 113}, 231605 (2014).
% ``Infinite class of PT-symmetric theories from one timelike Liouville
% Lagrangian'' [arXiv: hep-th/1408.2432]

\bibitem{R15} C.~M.~Bender, Rep.~Prog.~Phys.~{\bf 70}, 947 (2007).

\bibitem{R16} V.~Branchina, E.~Messina, and M.~Sher, Phys.~Rev.~D {\bf 91},
013003 (2015).
% "Lifetime of the electroweak vacuum and sensitivity to Planck scale physics"

\bibitem{R17} C.~M.~Bender, D.~D.~Holm, and D.~W.~Hook, J.~Phys.~A:
Math.~Theor.~{\bf 40}, F81 (2007);
% [arXiv: math-ph/0609068] F81-F89 "Complex Trajectories of a Simple Pendulum"
C.~M.~Bender, D.~C.~Brody, and D.~W.~Hook, J.~Phys.~A: Math.~Theor.~{\bf 41},
352003 (2008); 
% [arXiv: hep-th/0804.4169]
% "Quantum effects in classical systems having complex energy"
A.~Cavaglia, A.~Fring, and B.~Bagchi, J.~Phys.~A: Math.~Theor. {\bf 44}, 325201
(2011).
% " PT-symmetry breaking in complex nonlinear wave equations and their
% deformations"

\bibitem{R18} C.~M.~Bender and S.~A.~Orszag, {\it Advanced Mathematical
Methods for Scientists and Engineers} (McGraw Hill, New York, 1978).

\bibitem{R19} The spectrum of $H=p^2-ix$ is null; see I.~Herbst,
Commun.~Math.~Phys.~{\bf 64}, 279 (1979). This is because for this problem there
are precisely three Stokes wedges of angular opening $120^\circ$. If the
solution to the Schr\"odinger equation vanishes exponentially in one wedge, it
grows exponentially in the adjacent two wedges and thus no eigenvalue condition
is possible. The branch cut allows the Hamiltonian $H=p^2-ix\log(ix)$ to evade
this constraint; there are infinitely many Stokes wedges on the Riemann surface.

\bibitem{R20} C.~M.~Bender and P.~D.~Mannheim, Phys.~Lett.~A  {\bf 374}, 1616
(2010).
% "PT Symmetry and Necessary and Sufficient Conditions for the Reality of
% Energy Eigenvalues" arXiv:0902.1365 [hep-th]

\end{thebibliography}
\end{document}